\documentstyle[11pt]{article}

\columnwidth\textwidth

\def\kon#1#2{\vbox{\halign{##&&##\cr
\lower4pt\hbox{$\scriptscriptstyle\vert$}\hrulefill &
\hrulefill\lower4pt\hbox{$\scriptscriptstyle\vert$}\cr $#1$& $#2$\cr}}}

\title{Smeared and unsmeared chiral vertex operators} \author{Florin
Constantinescu \\Fachbereich Mathematik, \\ Johann Wolfgang Goethe Universit\"at
Frankfurt, \\ Robert-Mayer-Str. 10, D-60054 Frankfurt am Main, Germany \and
G\"unter Scharf \\ Institut f\"ur Theoretische Physik, \\ Universit\"at
Z\"urich, \\ Winterthurerstr. 190 , CH-8057 Z\"urich, Switzerland}

\date{} 
\begin{document} \maketitle \vskip 3cm \begin{abstract}  We prove
unboundedness and boundedness of the unsmeared and smeared chiral vertex
operators, respectively. We use elementary methods in bosonic Fock space, only.
Possible applications to conformal two - dimensional quantum field theory,
perturbation thereof, and to the perturbative construction of the sine-Gordon
model by the Epstein-Glaser method are discussed. From another point of view the
results of this paper can be looked at as a first step towards a Hilbert space
interpretation of vertex operator algebras.  \end{abstract} \newpage

\def\d{\partial}\def\=d{\,{\buildrel\rm def\over =}\,}
\def\dh{\mathop{\vphantom{\odot}\hbox{$\partial$}}} \def\dl{\dh^\leftrightarrow}
\def\sqr#1#2{{\vcenter{\vbox{\hrule height.#2pt\hbox{\vrule width.#2pt
height#1pt \kern#1pt \vrule width.#2pt}\hrule height.#2pt}}}}
\def\w{\mathchoice\sqr45\sqr45\sqr{2.1}3\sqr{1.5}3\,} \def\eps{\varepsilon}
\def\oe{\overline{\rm e}} \def\onu{\overline{\nu}}
\def\ds{\hbox{\rlap/$\partial$}} \def\psq{{\overline{\psi}}}
\def\la{{(\lambda)}} \def\lap{\bigtriangleup\,}

\section{Introduction}

The subject of massless two-dimensional fields was always a source of
interesting problems. The light-cone variables, when compactified in the
euclidean, allow application of complex methods. Here we are concerned, in the
compact case, with both unsmeared and smeared vertex operators near and on the
unit circle.  Using elementary methods only, we show that, as operators in
Hilbert spaces which are related to the bosonic Fock space, the usual unsmeared
vertices are poor operators, whereas the smeared ones are nice bounded
operators. The result is surprising taking into account that bosonic operators
are usually unbounded. On the other hand, two-dimensional abelean bosonization
makes the result plausible, at least in some case (equivalence to fermiones
which are bounded).

Although the functional properties of the unsmeared vertex operators are not
overwhelming their algebraic properties, when restricted inside the unit circle,
are remarkable. This is consistent with their usefullness in the frame of vertex
operator algebras.

In the smeared case we were motivated by similar results obtained in the
framework of the Wess-Zumino-Witten model of two-dimensional conformal quantum
field theory [1, 2] and for Minkowski two-dimensional massless fields [3] by
explicit fermionic methods. Instead we keep working in the bosonic Fock space
where vertex operators naturally live. Our main tools are a generalized Gram
inequality for determinants, an explicit tensor product argument which remembers
of a trick used in [1, 3] and a further extension of it.

The range of validity of our results extends to chiral vertex operators with
charges in the closed unit circle. We expect some input on two-dimensional
conformal quantum field theory and on the sine-Gordon model. Indeed, from the
conformal field theory point of view the latter is strongly related to the
perturbed conformal quantum field theory [20].  The consequences are twofold:
(i) On one side, in a hamiltonian approach to perturbed conformal field theory
the boundedness of the vertex operator (which appears as perturbation) would
suggest a regular analytic perturbation which is seldom even in quantum
mechanics. This agrees with a convergent perturbation series in the lagrangean
formalism [4]. (ii) On the other side, a perturbative construction of the
S-matrix for the sine-Gordon model by the Epstein-Glaser method [5, 6] appears
to provide us with a convergent perturbation series (before the adiabatic
limit).  Needless to say that, although our study in this paper is restricted to
the compact case, we expect similar results by similar methods in the
non-compact Minkowski case too; the case in which the Epstein-Glaser method is
currently used [5, 6]. We will return to this subject elsewhere. For the usual
perturbative approach including the case of conformal quantum field theory see
[7, 8, 9].

The paper is organized as follows. In the second section we set up the bosonic
Fock space notations, define the unsmeared chiral vertex operators and discuss
their Hilbert space properties. We follow here [10, 11, 12] with some
improvements. In the third section we introduce the smeared vertex operatores on
the unit circle in bosonic Fock space and prove some inequalities for their
vacuum expectation values in a special case. Here the (generalized) Gram
inequality, proven in Appendix 1, is used. Gram determinant inequalities are a
hint to fermions behind the bosons, but we prefer to stay in the bosonic
framework and in fact find, if possible, alternative proofs which are not
necessarily fermionic in nature.

In the forth and fifth section we adjust the bosonic Fock space framework in
order to incorporate neutrality and finally interpret the results of the third
section as boundedness of the smeared chiral vertex operator with neutrality
condition. We use a method which is standard in the old style approach to string
theory (and was taken up to conformal field theory and study of some infinite
dimensional Lie algebras). A possible alternative aproach is mentioned in
section 6. In that section a discussion of the results obtained and perspectives
concerning conformal field theory and sine-Gordon model follows. Appendices
provide results used in the main text, but they can also be of independent
interest.

\section{Unsmeared chiral vertex operators in bosonic Fock space}

Let $a_n$, $n\in {\bf Z}-\{0\}$ generate the Heisenberg algebra
$[a_n,a_m]=n\delta_{n,-m}$ where $a_n$, $n\ge 1$ are annihilation and $a_{-n}$,
$n\ge 1$ creation operators in bosonic Fock space $\cal F$. We will also
consider central extensions by $a_0$ with $a_0\Phi_0(\alpha) =
\alpha\Phi_0(\alpha)$, where $\Phi_0(\alpha)$ is the cyclic vacuum in the
bosonic Fock space $\cal F(\alpha)$, now indexed by the basic charge $\alpha$
(in particular we can take $\alpha=0$). As Hilbert spaces the $\cal F(\alpha)$
are all the same and we will keep denoting them by $\cal F$ with vacuum
$\Omega\equiv\Phi_0$ instead of $\Phi_0(\alpha)$ if clear from the context. In
$\cal F$ we consider the usual basis $$\Phi_\eta={1\over
(\eta!I^\eta)^{1/2}}\,a_{-k}^{\eta_k}\ldots a_{-1}^{\eta_1}\Phi_0,\eqno(1)$$
where $$\eta=(\eta_1,\eta_2,\ldots),\quad \eta_i\ge 0,\quad \eta!=\prod
_{i=1}^\infty\eta_i!$$ $$I^\eta=\prod_{i=0}^\infty i^{\eta_i},\quad
\|\eta\|=\sum_{i=1} ^\infty i\eta_i<\infty.$$

Let ${\cal F}_0\subset{\cal F}$ be the linear span of $\Phi_\eta$. The
(unbounded, closed, densely defined) operators $a_n$, $n\ne 0$ act as usual for
$n\ge 1$ $$a_n\Phi_\eta=\sqrt{n\eta_n}\Phi_{\eta-e_n}$$
$$a_{-n}\Phi_\eta=\sqrt{n(\eta_n+1)}\Phi_{\eta+e_n},\eqno(2)$$ where $e_n$ is
the unit vector in $l^2$ with zero components for $k\ne n$ and one for $k=n$.
Let $\gamma, z\in{\bf C}$, $z\ne 0$. The formal unsmeared vertex operator in
$\cal F$ is $$V_\gamma(z)\equiv V(\gamma,z)=V_-(\gamma,z)V_+(\gamma,z)\eqno(3)$$
with $$V_-(\gamma,z) =\exp\biggl(\gamma\sum_{n=1}^\infty{z^n\over n}a_{-n}
\biggl)$$ $$V_+(\gamma,z)=\exp\biggl(-\gamma\sum_{n=1}^\infty{z^{-n}\over n}a_n
\biggl).\eqno(4)$$ Further on we consider vertex operatots $\tilde
V_\gamma(z)\equiv \tilde V(\gamma,z)$ defined as follows. First consider the
operator $$T_\gamma:\quad {\cal F}(\alpha)\rightarrow {\cal F}(\alpha+\gamma)
\eqno(5)$$ such that $[T_\gamma, a_n]=0$ for $n\ne 0$ and $[T_\gamma, a_0]=
-\gamma T_\gamma$. One can check that this is a homomorphism of Heisenberg
moduls. We introduce the vertex operator $\tilde V(\gamma,z)$ from ${\cal
F}(\alpha)$ to ${\cal F}(\alpha+\gamma)$ by $$\tilde V_\gamma(z)\equiv\tilde
V(\gamma,z)=T_\gamma z^{\gamma a_0} V(\gamma,z).\eqno(6)$$ We will use both
vertices $V(\gamma,z)$ and $\tilde V(\gamma,z)$. Since the operators $T_\gamma$
and $a_0$ are harmless, there will be, from the Hilbert space point of view, not
much difference between the two vertices. Some difference will appear later
after introducing the neutrality condition.

Now let us introduce the involution $a_n^+=a_{-n}$, $n\in {\bf Z}$. The formal
adjoints of vertex operators are again vertex operators
$$V(\gamma,z)^+=V\Bigl(-\gamma^*,{1\over z^*}\Bigl)$$ $$\tilde
V(\gamma,z)^+=\tilde V\Bigl(-\gamma^*,{1\over z^*}\Bigl), \eqno(7)$$ where $z^*,
\gamma^*$ are  the complex conjugate of $z, \gamma\in{\bf C}$.  For the purpose
of computations to follow we remark that $[a_0,T_\gamma]=\gamma T_\gamma$
implies $$z^{\gamma_1a_0}T_{\gamma_2}=z^{\gamma_1\gamma_2}T_{\gamma_2}
z^{\gamma_1a_0}.$$ The proof is a simple computation.

Now we start looking at vertices no longer formal (as in (4)) but as operators
in Hilbert space. We restrict here to $V=V(\gamma,z)$ as operator in ${\cal
F}(\alpha)$ with $\alpha=0$ (denoted by $\cal F$) but similar results hold for
$\tilde V$, too. A direct computation in Fock space [10, 11] shows that we
obtain well-defined matrix elements of $V$:
$$v_{\eta,\nu}(\gamma,z)=(\Phi_\eta,V(\gamma,z)\Phi_\nu)=$$ $$={1\over\sqrt{\eta
!\nu !}}\prod_{i=1}^\infty m_{\eta_i\nu_i}\Bigl(
{\gamma\over\sqrt{i}}z^i,-{\gamma\over\sqrt{i}}z^{-i}\Bigl)\eqno(8)$$ where
$$m_{\eta\nu}(x,y)=\sum_{j=0}^{\min (\eta,\nu)}{\eta\choose j}{\nu\choose j}
j!x^{\eta-j}y^{\nu-j}\eqno(9)$$ are related to the monic Charlier and Laguerre
polynomials [13] $$C_n^{(a)}(x)=n!L_n^{(x-n)}(a)=\sum_{l=0}^n {n\choose
l}{x\choose l} l!(-a)^{n-l}\eqno(10)$$ by
$$m_{\eta\nu}(x,y)=y^{\nu-\eta}C_\eta^{-(xy)}(\nu).\eqno(11)$$ Note that the
product in (8) is finite because $m_{\eta_i\nu_i}$ is different from zero for
only a finite number of $i$ and $m_{00}(x,y)=1$. The generating function of
$C_n^{(a)}(x)$ is $$\sum_{n=0}^\infty C_n^{(a)}(x){w^n\over n!}=e^{-aw}(1+w)^x
\eqno(12)$$ and the orthogonality relation $$\int\limits_0^\infty
C_m^{(a)}(x)C_n^{(a)}(x)\,d\psi^{(a)}(x)=a^n n!  \delta_{m,n}\eqno(13)$$ holds
with respect to the step function $\psi^{(a)}$ with jumps $${e^{-a}a^x\over
x!},\quad\sum_{x=0}^\infty {e^{-a}a^x\over x!}=1 \eqno(14)$$ at
$x=0,1,2,\ldots$. For $a>0$ (our case if $\gamma$ real) this is the Poisson
distribution. At this point it is interesting to remark that the basic formula
for giving a Hilbert space meaning to formal products of vertex operators (see
later) corresponds to the following generalization of the orthogonality relation
for the Charlier polynomials [13]:

{\bf Lemma 1} For $x,y,z\in{\bf C}$ and $i,j$ integers $$\sum_{k=0}^\infty
{1\over k!}m_{ik}(x,y)m_{kj}(z,w)=m_{ij}(x+z,y+w) e^{xy}.\eqno(15)$$ The proof
which is a long induction argument can be found in [10].  It encodes the fact
that formally the product of two vertex operators is, up to scalar factors,
again a vertex-like operator.

Generally the matrix elements $v_{\eta\nu}=v_{\eta\nu}(\gamma,z)$ define an
operator $V=V(\gamma,z)$ in Hilbert space $\cal F$ (no longer formal) by
$$V\Psi=\sum_\eta\sum_\nu v_{\eta\nu}(\Phi_\nu,\Psi)\Phi_\eta\eqno(16)$$ for
$\Psi$ in the domain of definition $$D(V)=\Bigl\{\Psi\in{\cal
F}:\quad\lim_{k\to\infty}\sum_{\|\nu\|\le k} v_{\eta\nu}(\Phi_\nu,\Psi)\quad{\rm
exists}$$ $$\hbox{\rm for all $\eta$ and}\>\sum_\eta\Bigl\vert\sum_\nu
v_{\eta\nu} (\Phi_\nu,\Psi)\Bigl\vert^2<\infty\Bigl\}.\eqno(17)$$ Certainly the
domain of definition of $V$ can be void, $D(V)=\emptyset$.  In this case the
matrix elements $v_{\eta\nu}$ determine $V$ only as bilinear form and not
properly as an operator. This will really happen in some cases below.  Using
definition (16, 17), a bunch of results on $V=V(\gamma,z)$ has been proven [11]
using mainly lemma 1 and coherent states [14] generated by exponentials of type
$V_-$ in (4). We select what is relevant for us:

{\bf Theorem 2} We have for arbitrary $\gamma\in {\bf C}$

(i) For $|z|<1$, $V(\gamma,z)$ is densely defined with ${\cal F}_0$ in its
domain of definition $D(V(\gamma,z))$.

(ii) For $|z|>1$, $V(\gamma,z)$ is closed.

(iii) For $|z|<1$, $V(\gamma,z)$ is not closable.

(iv) Let $|z_2|<|z_1|<1$, then $$V(\gamma,z_2){\cal F}_0\subset
D(V(\gamma,z_1)).$$

We recall that a Hilbert space operator $T$ is closed if its graph $G(T)$ is
closed; $T$ is said to be closable if the closure $\overline{G(T)}$ is a graph.

Proofs of the theorem are based on lemma 1 and can be found in [10, 11]. In [11]
even stronger results are proven. In particular, more involved considerations
[11] enable one to strengthen property (ii) to

(ii') For $|z|>1$ the domain of definition of $V(\gamma,z)$ is void and as such
$V(\gamma,z)$ is trivially closed.

Taking into account definition (7) of the formal adjoint, the situation with
$V(z), V^+(z)$ appear to be somewhat similar to that of annihilation and
creation operators $a(x), a^+(x)$ in elementary quantum mechanics.  Indeed, the
domain of definition of $a^+(x)$ as operator in Fock space is void, but working
with bilinear forms instead of operators saves the matter. Only after smearing,
$a^+(f)$, $f\in L^2$ becomes a (nontrivial) operator.  However, the reader
should not push this analogy too far because of chiral properties of $V, V^+$
which are absent in $a, a^+$.  At this stage we retain the fact that there is a
striking asymmetry between unsmeared vertices inside and outside the unit
circle, as far as their operator properties in the Fock space ${\cal F}$ are
concerned. The symmetry is restored after smearing on the unit circle as we will
see below.

According to property (ii) the (maximal) operator $V(\gamma,z)$ defined by (16)
(17) is not closable. It could be that a restriction of $V(\gamma,z)$ is
closable although $V(\gamma,z)$ itself doesn't have this property. The assertion
about the non-closability of $V(\gamma,z)$ can be extended as follows [11]: Let
us define the "minimal" vertex operator as
$$V(\gamma,z)_0=V(\gamma,z)\Bigl\vert_{{\cal F}_0}$$ with
$D(V(\gamma,z)_0)={\cal F}_0$. It is clear that each extension of
$V(\gamma,z)_0$ (in particular $V(\gamma,z)$) is not closable if $V(\gamma,z)_0$
is not closable. The following property [11] is a generalization of (iii):

(iii') For $|z|<1$, $V(\gamma,z)_0$ is not closable.

{\bf Corollary 3} For arbitrary $\gamma\in {\bf C}$ and $|z|<1$ the vertex
$V(\gamma,z)$ is an unbounded operator in Fock space.

This is a consequence of property (iii). We give an independent simple proof
\footnote{We thank W.Boenkost for suggesting this proof to us.} (cf. [12]). Let
us define for $z,\zeta\in {\bf C}$, $l\in {\bf N}$ coherent states as
$$|z,\zeta,l\rangle =N_l\exp\Bigl(\zeta\sum_{n=1}^l{z^n\over n}a_{-n}
\Bigl)\Phi_0,\eqno(18)$$ where Dirac's notation has been used and $N_l$ is the
normalization constant $$N_l=\exp\biggl(-{|\zeta|^2\over
2}\sum_{n=1}^l{|z|^{2n}\over n}\biggl).\eqno(19)$$ The coherent state (18) can
be expanded as $$|z,\zeta,l\rangle =N_l\sum_{\eta_1,\ldots,\eta_l=0}^\infty\prod
_{n=1}^l{z^{n\eta_n}\zeta^{\eta_n}\over (\eta_n!n^{\eta_n})^{1/2}}
\,\Phi_\eta.\eqno(20)$$ The limit of (20) for $l\to\infty$ exists in Fock space
$$|z,\zeta\rangle =\lim_{l\to\infty}|z,\zeta,l\rangle.\eqno(21)$$ On the other
hand the coherent states are eigenvectors of annihilation operators $a_n$, $n\ge
0$ $$a_n|z,\zeta,l\rangle=\cases{\zeta z^n|z,\zeta,l\rangle, & for $n\le l$\cr 0
&for $n>l$\cr}.\eqno(22)$$ Using the fact that such $a_n$ are closed operators,
it follows $$a_n|z,\zeta\rangle=\zeta z^n|z,\zeta\rangle.\eqno(23)$$ These
properties are used to compute
$$\langle\Phi_0|V(\gamma,z)|z,\zeta,l\rangle=\exp\Bigl(-\gamma\zeta
\sum_{n=1}^l{1\over n}\Bigl)\langle\Phi_0|z,\zeta,l\rangle$$
$$=N_l\exp\Bigl(-\gamma\zeta\sum_{n=1}^l{1\over n}\Bigl).\eqno(24)$$ It follows
for $\zeta=-\gamma^*$ that $$\langle\Phi_0|V(\gamma,z)|z,-\gamma^*,l\rangle=
N_l\exp\Bigl(|\gamma|^2\sum_{n=1}^l{1\over n}\Bigl).\eqno(25)$$ This shows that
the matrix elements of $V(\gamma,z)$ diverge for $l\to\infty$ and the
unboundedness follows.

Properties (i)-(iii) say nothing about the case $|z|=1$.  Property (iii) shows
that $V(\gamma,z)$ in the interesting region $|z|<1$ is a poor operator.
Nevertheless, Property (iv) allows for defining products
$$V(\gamma_1,z_1)V(\gamma_2,z_2)\ldots V(\gamma_r,z_r)\eqno(26)$$ for
$|z_r|<|z_{r-1}|<\ldots <|z_1|<1$ as densely defined operatots in $\cal F$ with
${\cal F}_0$ in their domains.

Some remarks are in order:

First we didn't use any kind of braid relation between vertices and in fact
defined Hilbert space products $V(\gamma_1,z_1)\ldots V(\gamma_r,z_r)$ only for
$|z_r|<\ldots <|z_1|<1$.

Second we didn't introduce neutrality condition, common in massless
two-dimensional field theory. Certainly it is possible to introduce the braiding
relation consistent with the properties of unsmeared $V(\gamma,z)$ but it
doesn't help too much for improving  their Hilbert space properties (i) - (iv).
The situation is different in the smeared case as we will see later. We didn't
try to find out if neutrality could improve the properties of unsmeared
vertices.

Third, it is interesting to remark that the product of vertices (26) is defined
in ${\cal F}$ without an invariant domain for the factors.  Indeed, ${\cal F}_0$
is not invariant under $V(\gamma,z)$. This is not the situation one is used to
have in quantum field theory where the invariance of the domain on which the
field operators are densely defined is part of the axioms. Later on in this
paper the situation will change by passing to some modifications of ${\cal F}$.

This is the situation with unsmeared vertex operators. In the next section we
will show that a smearing operation applied to vertices dramatically improves
their properties (near $|z|=1$) such that finally under the neutrality condition
they turn into bounded operators in the bosonic Fock space to be precisely
defined below. We consider the case $|\gamma|\le 1$ in this paper but there are
indications that the result could be extended.

For later use we mention the following formula which now has a Hilbert space
operator interpretation (see the remarks above concerning the existence of
products) $$\tilde V(\gamma_1,z_1)\tilde V(\gamma_2,z_2)\ldots\tilde
V(\gamma_r,z_r) =\prod_{1\le i<j\le r}\Bigl(1-{z_j\over
z_i}\Bigl)^{\gamma_i\gamma_j} T_{\sum\gamma_j}\times$$
$$\times\prod_{i=1}^rz_i^{\gamma_ia_0}\exp\biggl(\sum_{n=1}^\infty {1\over
n}(\sum_{i=1}^r\gamma_iz_i^n)a_{-n}\biggl)\exp\biggl(\sum_{n=1} ^\infty{1\over
n}(\sum_{i=1}^r\gamma_iz_i^{-n})a_n\biggl)\eqno(27)$$ for
$|z_r|<|z_{r-1}|<\ldots <|z_1|<1$. In the case of $V(\gamma_i,z_i)$ obvious
factors in (27) have to be left out. In particular (27) reproduces the well
known formula for the $n$-point function under neutrality $\sum\gamma_i=0$ (here
$\alpha=0$) $$(\Phi_0,\tilde V(\gamma_1,z_1)\ldots\tilde
V(\gamma_n,z_n)\Phi_0)=\prod_{i<j} (z_i-z_j)^{\gamma_i\gamma_j}.\eqno(28)$$ The
determination in (27) and (28) is fixed as usual by taking $\log (z_i-z_j)$ real
for $0<z_j<z_i$.

\section{Smeared chiral vertex operators}

The formal smeared chiral vertex operator on the unit circle $S^1$ is
$$V(\gamma,f)={1\over 2\pi}\int\limits_{³z³=1}V(\gamma,z)f(z)dz\eqno(29)$$ where
$f=f(z)$ is a test function on $S^1$ to be chosen below. In this section we
start by looking at (29) as a limit for $z$ approaching the unit circle from the
interior, i.e. (29) has to be understood as $$V(\gamma,f)=\lim_{r\to 1-}{1\over
2\pi}\int\limits_{|z|=1} V(\gamma,rz)f(z)dz. \eqno(30)$$ Let us consider the
case $\gamma=1$ first. When appropriate we use the notation
$V(\gamma=1,z)=V(z)$. A rough idea of what happens is obtained by calculating
the following norm in the unsmeared case for $|z|=1-\eps$, $\eps\to 0+$ with
help of (27) $$\|V(z)\Phi_0\|^2=(\Phi_0,V(z)^+V(z)\Phi_0)=$$ $$={1\over
1-z^*z}\Bigl\vert_{|z|=1-\eps,\eps\to 0+}=\infty,\eqno(31)$$ whereas in the
smeared case with $f\in L^2(S^1)$
$$\|V(f)\Phi_0\|^2=(\Phi_0,V(f)^+V(f)\Phi_0)=$$ $$={1\over
4\pi^2}\int\int\limits_{|z|=|w^*|=1}{1\over 1-w^*z} f^*(w^*)f(z)\,dw^*\,
dz=\sum_{n=1}^\infty |c_{-n}|^2\le \|f\|_2^2<\infty.\eqno(32)$$ Here $c_n$ is
the Fourier coefficient $$c_n={1\over 2\pi
i}\int\limits_{|z|=1}z^{-n-1}f(z)\,dz= {1\over
2\pi}\int\limits_0^{2\pi}e^{-in\theta}f\Bigl(e^{i\theta}\Bigl)\,
d\theta\eqno(33)$$ and the $L_2(S^1)$ norm is given by $$\| f\|_2^2={1\over
2\pi}\int\limits_0^{2\pi}|f(e^{i\theta})|^2\,
d\theta=\sum_{n=-\infty}^{+\infty}|c_n|^2.$$ In (32) use was made of the formal
smeared adjoint $$V(\gamma,f)^+={1\over
2\pi}\int\limits_{|z|=1}V(\gamma,z)^+f^*(z) \,dz^*={1\over 2\pi}\int\limits
_{|z|=1}V\Bigl(-\gamma^*,{1\over z^*}\Bigl)f(z^*)^*\,dz^*,$$ Following the
convention in (31), $z$ approaches the unit circle from the interior and
consequently $1/z^*$ approaches it from the exterior. In (32), before taking the
limit to the unit circle, we have $|z|<1<|w^*|^{-1}$ which assures the validity
of the geometric series used there. The rigorous definition of the smeared
vertex $V(\gamma,f)$ and adjoint $V(\gamma,f)^+$ as operators in bosonic Fock
space is analogous to the definition of their unsmeared counterparts in section
2 (cf. further in this section). In addition we have to remark that the smearing
in $L^2$ enables us to take them on the unit circle, a fact which is not true in
the unsmeared case. This remembers the computation in (32) as opposed to (31).
For the convenience of the reader let us give some details in the special case
$V(\gamma,f)\Phi_0$. Indeed, using formulas from section 2 we have
$$\lambda_\eta\equiv\lambda_\eta(z)=(\Phi_\eta, V(\gamma,z)\Phi_0)=
\prod_{i=1}^\infty (\eta_i!)^{-1/2}\biggl({\gamma z^i\over\sqrt{i}}
\biggl)^{\eta_i}$$ and $$\sum_\eta
|\lambda_\eta|^2=\sum_\eta\prod_{i=1}^\infty{1\over\eta_i!} \biggl({|\gamma
z^i|^2\over i}\biggl)^{\eta_i}= $$
$$=\prod_{i=1}^\infty\biggl(\sum_{k=0}^\infty{1\over k!} \biggl({|\gamma
z^i|^2\over i}\biggl)^k\biggl)=$$
$$=\prod_{i=1}^\infty\exp\biggl({|\gamma|^2|z^{2i}|\over i}\biggl)=
\exp\biggl(|\gamma|^2\sum_{i=1}^\infty{|z^2|^i\over i}\biggl)=$$
$$=\Bigl({1\over 1-|z|^2}\Bigl)^{|\gamma|^2},$$ showing that in the unsmeared
case $\sum_\eta |\lambda_\eta|^2<\infty$ for $|z|<1$, but not for $|z|=1$
because of the divergence of the harmonic series. On the other hand, in the
smeared out case for $$\Lambda_\eta={1\over
2\pi}\int\limits_{|z|=1}\lambda_\eta(z)f(z)\,dz\eqno(34)$$ we have
$\sum_\eta|\Lambda_\eta|^2<\infty$ as a consequence of the computation above and
Parseval's relation for the Fourier coefficients of $f\in L^2(S^1)$. This shows
that the vacuum is in the domain of definition of $V(\gamma,f)$ when
concentrating on the unit circle which was not the case for the unsmeared $V$
(see (31)). Similar considerations hold for products of smeared vertex operators
appearing below. We leave the details to the interested reader.

We remark that some care is needed for the case $|z|>1$ (including the adjoint
with $|z|<1$). The truncation which makes divergent sums well defined is
provided by looking at the problem in the framework of bilinear forms. The way
in which the bilinear adjoint operation is implemented is different from the
standard case of annihilation and creation operators. It can be understood in
terms of Fourier coefficients of the smearing function: under adjunction they
reverse the index. On products of smeared vertices applied to the vacuum the
adjoint operation is equivalent (according to the Hardy decomposition) to the
change of the regularization in the standard kernel from inside to the outside
of the unit circle. In this way analytic continuation and braiding enter the
scene. Consequently the results are fully consistent with those obtained by
usual formal work supplemented by analytic continuation, braiding etc. At the
same time, for this particular example, one obtains a Hilbert space version of
methods in operator vertex algebras (formal fields, formal distributions etc.).

Anticipating, the smearing operation reinforces symmetry: both $V(f)$, $f\in
L^2$ and its adjoint are densely defined, closed operators.  Neutrality will
turn them into bounded operators.  We didn't study further $V(f)$, $f\in L^2$
without neutrality condition here.

Remark that even for the case $|\gamma|\le 1$ the norm $\|V(\gamma,f)\Phi_0\|^2$
is finite for $f\in L^2(S^1)$. Indeed, for $1>x=|\gamma|^2>0$ the binomial
series $(|a|<1)$ $$(1-a)^{-x}=1+{x\over 1!}a+{x(x+1)\over
2!}a^2+{x(x+1)(x+2)\over 3!} a^3+\ldots$$ gives
$$\|V(\gamma,f)\Phi_0\|^2={x\over 1!}|c_{-1}|^2+{x(x+1)\over 2!}
|c_{-2}|^2+\ldots\le$$ $$\le {1\over 1!}|c_{-1}|^2+{1\cdot 2\over
2!}|c_{-2}|^2+\ldots =\|V(\gamma=1,f)\Phi_0\|^2<\infty.\eqno(35)$$ The two-point
function results (32) and (35) in the smeared out case are encouraging as oposed
to the unsmeared case (31). The estimates by geometric series expasion above can
be replaced by the more efficient technique of Hardy spaces (see for instance
[15]). In this theory we have the direct sum decomposition
$$L^2(S^1)=H_+^2\oplus H_-^2\eqno(36)$$ where $H_+^2$ and $H_-^2$ are also
Hilbert spaces of $L^2$-functions with positive and zero frequencies and
negative frequencies, respectively. $H_+^2$ is the usual Hardy space denoted by
$H^2$. In a different language we have in $H_+^2$ $L^2$-boundary values from
inside and in $H_-^2$ from outside the unit circle. The main formula we use in
this context is $${1\over 2\pi i}\int\limits_{S^1}\int\limits_{S^1}{1\over
z_2-z_1} f(z_2)g(z_1)\, dz_2\,dz_1=$$
$$=\int\limits_{S^1}f^{(+)}(z_1)g(z_1)\,dz_1=
\int\limits_{S^1}f^{(+)}(z_1)g^{(+)}(z_1)\,dz_1,\eqno(37)$$ where the
integration variables tend to the unit circle, respecting $|z_2|>1>|z_1|$. We
use (37) in several forms which at first glance look different but are always
the same formula (37). For instance, we have $$(V(f)\Phi_0, V(g)\Phi_0)=(\Phi_0,
V^+(f)V(g)\Phi_0)=$$ $$={1\over 4\pi^2}\int\limits_{|w^*|=|z|=1}{1\over
1-w^*z}f^*(w^*) g(z)\,dw^*\,dz={1\over 4\pi^2}\int{u^{-1}f^*(u^{-1})g(z)\over
u-z}\,du \,dz=$$ $$={1\over 2\pi
i}\int\Bigl(z^{-1}f^*(z^{-1})\Bigl)^{(+)}g(z)\,dz =(f^{(-)},g^{(-)}).\eqno(38)$$
From (38) we get in particular for $f=g$ the previous relation (32) from which
we retain $$\|V(f)\Phi_0\|=\|f^{(-)}\|_2\le\|f\|_2.\eqno(39)$$

In the following we will generalize the relation (39) to scalar products of the
form $$(V_n(f),V_n(g)),\eqno(40)$$ where $$V_n(f)=V(f_1)V(f_2)\ldots
V(f_n)\Phi_0\eqno(41)$$ and similarly for $V_n(g)$, with $f_i(z_i), g_i(z_i)\in
L^2(S^1)$, $i=1,2,\ldots ,n$ and the regularization prescription
$$|z_n|<|z_{n-1}|<\ldots <|z_1|<1.$$ We write for $f=(f_1,f_2,\ldots ,f_n)$,
$g=(g_1,g_2,\ldots ,g_n)$ $$(V_n(f),V_n(g))=(\Phi_0,V^+(f_n)\ldots
V^+(f_1)V(g_1)\ldots V(g_n) \Phi_0)=$$ $$={1\over (2\pi)^n}\int
D(w^*,z)\prod_{i=1}^n \underline{f}_i^*(w_i^*)\underline{g}_i(z_i)\,dw_i^*
\,dz_i\eqno(42)$$ where $$D(w^*,z)={\prod_{i<j}^n(z_i-z_j)(w_i^*- w_j^*)
\over\prod_{i,j=1}^n(1-z_i w_j^*)}\eqno(43)$$ and
$$\underline{f}_i(z_i)=z_i^{i-n}f_i(z_i)$$
$$\underline{g}_i(z_i)=z_i^{i-n}g_i(z_i)\eqno(44)$$ for $i=1,2,\ldots ,n$.

Using the Cauchy determinant formula $$D(w^*,z)=\det\Bigl({1\over
1-z_iw^*_j}\Bigl)_{1\le i<j\le n}\eqno(45)$$ and expanding the determinant we
get from (42) $$(V_n(f), V_n(g))=G_n(\underline{f}^{(-)};\underline{g}^{(-)})
\eqno(46)$$ where $$G_n(f,g)\equiv G_n(f_1,f_2,\ldots ,f_n;g_1,g_2,\ldots
,g_n)$$ $$=\pmatrix{(f_1,g_1)&\ldots&(f_1,g_n)\cr \ldots\cr (f_n,g_1)&\ldots&
(f_n,g_n)\cr}\eqno(47)$$ is the (generalized) Gram determinant.  For $f=g$ we
write for the usual Gram determinant $$G_n(f;f)\equiv G_n(f)\equiv
G_n(f_1,f_2,\ldots ,f_n).\eqno(48)$$ Similar relations hold for the vertices
$\tilde V(z)$. As a consequence of the smearing the Cauchy determinant which
often appears in two-dimensional (massless) physics (for instance in (28) with
$\gamma_i =\pm 1$ and neutrality $\sum\gamma_i=0$) goes over into a Gram
determinant. Certainly $G_n(f_1,\ldots ,f_n)\ge 0$, as necessary because from
(46) $$0<\|V_n(f)\|^2=G_n(\underline{f}).\eqno(49)$$ Observe that
$\underline{f}_i, \underline{g}_i\in L^2(S^1)$ iff $f_i,g_i\in L^2(S^1)$ and
$\|\underline{f}_i\|_2=\|f_i\|_2$, $\|\underline{g}_i\|_2=\|g_i\|_2$,
$i=1,2,\ldots ,n$. Using a simple Gram inequality (see Appendix 1 for a
collection of Gram determinant inequalities used in this paper) we get
$$\|V(f_0)V_n(f)\|\le\|\underline{f}_0^{(-)}\|_2\|V_n(f)\|\le
\|f_0\|_2\|V_n(f)\|.\eqno(50)$$ The two-point function estimate (38) over only
negative frequencies cannot be saved here because of the $z_i$-powers in (44).
The inequality (50) is a generalization of (39). Similar considerations apply to
$\tilde V$ with the same bounds $$\|\tilde V(f_0)\tilde
V_n(f)\|\le\|f_0\|_2\|\tilde V_n(f)\|,\eqno(51)$$ where $\tilde V_n$ is defined
as in (41) with $V$ replaced by $\tilde V$.  On the l.h.s. of (51) the norm is
taken in ${\cal F}(\alpha+n+1)$ (we have $\gamma=1$), whereas on the r.h.s. it
is taken in ${\cal F}(\alpha+n)$ with $\alpha$ arbitrary (in particular
$\alpha=0$).  In fact, here we can do better: for $\tilde V(f)$ there are no
powers of $z$ which force $f$ into $\underline{f}$ as in (44) and the bound is
$\| f^{(-)}\|_2$ involving only the negative frequency part in the Hardy
decomposition of $f$. This remark applies to all operators $\tilde V(f)$ to
follow. Indeed, instead of (42) we now have $$(\tilde V_n(f),\tilde
V_n(g))=(\tilde V(f_1)\ldots\tilde V(f_n) \Phi_0,\tilde V(g_1)\ldots\tilde
V(g_n)\Phi_0)=$$ $$={1\over (2\pi)^n}\int D(w^*,z)\prod_{i=1}^n
f_i^*(w_i^*)g_i(z_i)\,dw_i^*\,dz_i.$$ This formula is obtained by commuting
first the operator $T_\gamma$ through $a_0$ in $\tilde V_n(f)$ and $\tilde
V_n(g)$ separately and then using the adjoint operation on $V$'s as in (42). The
above formula proves the claim.

Let us pause for a moment in order to understand what we have achieved and what
we are going to do. First observe that up to now the elementary Gram determinant
inequality was used to estimate the norms (39) and (50) directly in the bosonic
Fock space, without appealing to any kind of abstract bosonic-fermionic
equivalence and without introducing the braiding relation. Our goal in the next
section is to prove boundedness of vertex operators in bosonic Fock space. We
will start working with the vertex operator $\tilde V(f)$ restricted from
$${\cal F}^\oplus=\oplus_{n=0}^\infty {\cal F}(\alpha+n)\quad {\rm to}$$
$$\tilde{\cal F}^\oplus=\oplus_{n=0}^\infty\tilde{\cal F}(\alpha+n), \eqno(52)$$
where $\alpha$ is arbitrary and $\tilde{\cal F}(\alpha+n)$ is the closed
subspace of ${\cal F}(\alpha+n)$ generated by $\tilde V_n(f)$ as in (41) i.e.
$$\tilde V_n(f)=\tilde V(f_1)\tilde V(f_2)\ldots\tilde V(f_n) \Phi(\alpha)$$
with $\Phi(\alpha)$ being the vacuum in ${\cal F}(\alpha)$ (see also [24]). In
chosing {\it orthogonal} direct sums in (52), the physical neutrality (the
"neutrality condition") is realized in both ${\cal F}^\oplus$ and $\tilde{\cal
F}^\oplus $. In physics neutrality is a consequence of the massless limit in
two-dimensional quantum field theory [18, 19]. This idea fits well in our
framework but we do not continue to discuss it. Let us only remark that in
quantum field theory the massless limit in each adequate framework requires
supplementary conditions (see [23] for instance) of which kind in our case the
neutrality is. Certainly, behind $\tilde{\cal F}^\oplus$ a fermionic structure
is hidden and this is related to the fact that on $V(f)$ (or $\tilde V(f)$) a
braiding relation can consistently be imposed, which in the case $\gamma=1$
degenerates into antisymmetry. This will no longer be the case in section 4
where $\gamma\ne 1$. Nevertheless it is exactly this "fermionic flavor" in
bosonic Fock space which does the job for us.

For proving boundedness of $\tilde V(f)$ in this framework we need a
(generalized) Gram determinant inequality to be proven in Appendix 1. In the
fifth section we will extend the results from the present $\gamma=1$ to
$\gamma\in [-1,1]$ and finally to $|\gamma|\le 1$. The Hilbert space
$\tilde{\cal F}^\oplus$ in (52) is a reasonable framework to study the vertex
$\tilde V(f)$ with neutrality condition. The vertex operator $V(f)$, on the
other hand, lives in the original bosonic Fock space ${\cal F}\equiv {\cal
F}(\alpha=0)$. This case, which is of interest too, is touched in section 5
where also the connection to the massless two-dimensional physics is shortly
discussed.

\section{Boundedness of vertex operators for $\gamma=1$ through generalized Gram
inequality}

By construction the set $$\tilde V_n(f)=\prod_{i=1}^n\tilde V(f_i)\Phi_0$$
$f=(f_1,f_2,\ldots ,f_n)$, $f_i\in L^2(S^1)$, $i=1,\ldots ,n$ and
$n=0,1,2,\ldots $ is total in $\tilde{\cal F}^\oplus$. We look first at
$$\Bigl\|\tilde V(f_0)\sum_{j=1}^m\alpha_j\tilde V_m(f_j)\Bigl\|,\eqno(53)$$
where $\alpha_j$ are complex constants. We changed the notation a little and
denoted now by $f_j$ in $\tilde V(f_i)$ the $n$-tuple
$f_j=(f_{j_1},\ldots,f_{j_n})$, $j=1,2,\ldots,n$. The norm in (53) is in ${\cal
F}(\alpha+n)$. To estimate this we use the following (generalized) Gram
inequality (see Appendix 1)
$$\sum_{i,j=1}^m\alpha_i^*\alpha_jG(f_0,f_i;f_0,f_j)\le\|f_0\|_2^2
\sum_{i,j=1}^m\alpha_i^*\alpha_jG(f_i;f_j).\eqno(54)$$ By expanding (53) and
using (54) as in (50) we get $$\Bigl\|\tilde V(f_0)\sum_{j=1}^m\alpha_j\tilde
V_n(f_j)\Bigl\|= \Bigl\|\sum_{j=1}^m\alpha_j\tilde V_{n+1}(f'_j)\Bigl\|\le$$
$$\le\|f_0\|_2\Bigl\|\sum_{j=1}^m\alpha_j\tilde V_n(f_j)\Bigl\|,\eqno(55)$$
where $f'_j=(f_0,f_j)$, $j=1,2,\ldots,m$ and in (55) obvious norms in ${\cal
F}(\alpha+n+1)$ and ${\cal F}(\alpha+n)$, respectively, are chosen. Taking into
account the direct sum structure of $\tilde{\cal F}^\oplus$ and the simple
operation (6) of $\tilde V$ in it, it follows that $\tilde V(f_0)$, $f_0\in
L^2(S^1)$ is bounded on a dense set and therefore bounded as operator in the
whole $\tilde{\cal F}^\oplus$ with bound $\|f_0\|_2$. Note that the braiding
condition on vertices (which would turn them into chiral fermions) was not
imposed. In fact vertices $\tilde V(z)$ can be explicitely realized as (chiral)
fermions in $\tilde {\cal F}^\oplus$.

\section{Boundedness of vertex operators with $|\gamma|\le 1$}

In this section we study vertices $\tilde V(\gamma,z)\equiv\tilde V_ \gamma(z)$
with $|\gamma|\le 1$ as operators in $${\cal
F}_\gamma^\oplus=\oplus_{n=0}^\infty {\cal F}(\alpha+n\gamma)$$ where $\tilde
V_\gamma(z)$ maps from ${\cal F}(\alpha)$ to ${\cal F}(\alpha+\gamma)$ in order
to account for the neutrality condition.  We restrict $\tilde V_\gamma(z)$ to
$$\tilde {\cal F}_\gamma^\oplus=\oplus_{n=0}^\infty\tilde {\cal F}
(\alpha+n\gamma).\eqno(56)$$ where $\tilde{\cal F}(\alpha+n\gamma)$ is the
 closed subspace of ${\cal F}(\alpha+n\gamma)$ generated by
$$\prod_{i=1}^n\tilde V_\gamma(f_i)\Phi_0,\quad f_i\in L^2(S^1),\quad
i=1,2,\ldots,n$$ and $n=0,1,2,\ldots$. Here $\tilde F(\alpha)$ is
one-dimensional generated by $\Phi_0(\alpha)=\Phi_0$.

We start with the case $\gamma\in [-1,1]$ and use a trick from [1, 3] which in
our bosonic framework becomes fully transparent. Consider two copies of
Heisenberg algebras generated by $a_n, a'_n$, $n\in {\bf Z}$ satisfying
$$[a_n,a_m]=n\delta_{n,-m}$$ $$[a'_n,a'_m]=n\delta_{n,-m}\eqno(57)$$
$$[a_n,a'_m]=0$$ and vertex operators $\tilde V_\gamma(z)$, $\tilde
V'_{\gamma'}(z)$ $$\tilde V_\gamma(z)=T_\gamma z^{\gamma
a_0}\exp\biggl(\gamma\sum_{n=1} ^\infty{z^n\over
n}a_{-n}\biggl)\exp\biggl(-\gamma\sum_{n=1}^\infty {z^{-n}\over n}a_n\biggl)$$
$$\tilde V'_{\gamma'}(z)=T'_{\gamma'}z^{\gamma a'_0}\exp\biggl(\gamma'
\sum_{n=1}^\infty{z^n\over n}a'_{-n}\biggl)\exp\biggl(-\gamma'\sum_{n=1} ^\infty
{z^{-n}\over n}a'_n\biggl)\eqno(58)$$ with
$$[T'_{\gamma'},a_0]=[T_{\gamma},a'_0]=[T_{\gamma},T'_\gamma]=0$$
$$[T_{\gamma},a_0]=-\gamma T_\gamma,\quad [T'_{\gamma'},a'_0]=-\gamma'
T'_{\gamma'}.\eqno(59)$$ Note that $\tilde V_\gamma(z)$, $\tilde
V'_{\gamma'}(z)$ refer to the same complex variable $z$ but act in different
Fock spaces. In the tensor product space we consider the tensor product operator
on the same variable $z$ $$\tilde V(z)=\tilde V_\gamma(z)\otimes\tilde
V_{\gamma'}(z)\eqno(60)$$ with $\gamma, \gamma'\in [-1,1]$,
$\gamma^2+\gamma^{\prime 2}=1$.  Such tensor products with a restriction to the
diagonal in the space-time variables were introduced long ago in axiomatic field
theory [26].  Observe that the notation in (60) which suggests a vertex $\tilde
V_{\gamma=1}(z)$ is consistent. Indeed, with $$A_n=\gamma a_n+\gamma'a'_n,\quad
n\in{\bf Z}$$ $$T=T_\gamma T'_{\gamma'}\eqno(61)$$ we have $$[A_n,A_m]=[\gamma
a_n+\gamma'a'_n,\gamma a_m+\gamma'a'_m]=n\delta_ {n,-m}(\gamma^2+\gamma^{\prime
2})=n\delta_{n,-m}\eqno(62)$$ and $$[T,A_0]=-(\gamma^2+\gamma^{\prime 2})T=-T$$
$$[T,A_n]=0,\quad n\in {\bf Z}-\{0\}.$$ We appologize for the sloppy
notation.One should read in (61) $$A_n=(\gamma a_n\otimes I)+(\gamma'a'_n\otimes
I')$$ where $I,I'$ are identities etc.  The involution $A_n^+=A_{-n}$ forces
$\gamma,\gamma'$ to be real. The tensor product $\tilde V(z)$ in (60) is an
extension of the vertex operator $$\tilde
V_{\gamma=1}(z)=Tz^{A_0}\exp\biggl(\sum_{n=1} ^\infty{z^n\over
n}A_{-n}\biggl)\exp\biggl(-\sum_{n=1}^\infty {z^{-n}\over
n}A_n\biggl)\eqno(63)$$ from its Fock space $\tilde F_{\gamma=1}^\oplus$ to the
tensor product of Fock spaces $\tilde F_{\gamma}^\oplus$, $\tilde F_{\gamma'}
^\oplus$, in which $\tilde V_\gamma(z)$, $\tilde V_{\gamma'}(z)$ live.

As in section 4, by the generalized Gram inequality the smeared vertex operator
$\tilde V_{\gamma=1}(f)$, $f\in L^2(S^1)$ is bounded in $\tilde{\cal
F}_{\gamma=1}^\oplus$. We claim that its smeared extension $\tilde V(f)$ in (60)
is also bounded with the same bound $\|f\|_2$ as operator in $\tilde{\cal
F}_{\gamma'}^\oplus\otimes \tilde{\cal F}_\gamma^\oplus$. This follows by
explicitly realizing $\tilde V(f)$ as a (chiral) fermionic operator satisfying
CAR on a dense domain (see [25]) not only in $\tilde {\cal F}^\oplus$ but also
in $\tilde{\cal F}_{\gamma'}^\oplus\otimes \tilde{\cal F}_\gamma^\oplus$ (cf.
[3]) and using the fact that the fermionic bound is algebraically determined
(see for instance theorem 1 in [25]). In fact looking at the proof of the
generalized Gram inequality in Appendix 1 one realizes that it is shaped after
the proof of theorem 1 in [25]. The only point is that in the Gram determinant
framework no braid relation (here responsable for antisymmetry) is necessary,
whereas the work with genuine fermions makes its introduction necessary (see
section 6 for a further remark on this point).

Let $\Psi,\Phi\in\tilde{\cal F}_\gamma^\oplus$ and $\Psi',\Phi'\in \tilde{\cal
F}_{\gamma'}^\oplus$. Then, for $f\in L^2(S^1)$ we can write
$$\Bigl(\Psi\otimes\Psi',\tilde V(f)(\Phi\otimes\Phi')\Bigl)={1\over 2\pi} \int
(\Psi,\tilde V_\gamma(z)\Phi)(\Psi',\tilde V_{\gamma'}(z)\Phi')
f(z)\,dz.\eqno(64)$$ We choose $\Phi'=\Omega'$ (vacuum $\Phi'_0(\alpha)$ in
$\tilde{\cal F} _{\gamma'}^{\oplus}$ now denoted by $\Omega'$) and $\Psi'=\tilde
V _{\gamma'}(g)\Omega'$ with $g(w)=w^{-1}$, $w\in S^1$. We write $$(\Psi',\tilde
V_{\gamma'}(z)\Phi')= (\tilde V_{\gamma'}(g)\Omega',\tilde
V_{\gamma'}(z)\Omega')=$$ $$=\biggl({1\over 2\pi}\int \tilde
V_{\gamma'}(w)g(w)\,dw\,\Omega', \tilde V_{\gamma'}(z)\Omega')\biggl)=$$
$$={1\over 2\pi}\int\limits_{|w^*|=1}\Bigl({1\over 1-w^*z}
\Bigl)^{\gamma^{\prime 2}} {1\over w^*}\,dw^*=i.\eqno(65)$$ Then
$$\Bigl(\Psi\otimes\Psi',\tilde V(f)(\Phi\otimes\Phi')\Bigl)= {i\over
2\pi}\int(\Psi,\tilde V_\gamma(z)\Phi)f(z)\,dz.\eqno(66)$$ Take $\gamma^{\prime
2}=1-\gamma^2$ and use the boundedness of $\tilde V(f)$ as operator in
$\tilde{\cal F}_{\gamma'}^\oplus\otimes \tilde{\cal F}_\gamma^\oplus$ to obtain
$$|(\Psi,\tilde V_\gamma(f)\Phi)|\le\|\Psi\|\|\Phi\|\tilde V_
{\gamma'}(g)\Omega'\|\|f\|_2.\eqno(67)$$ Now we use (35) to write $$\|\tilde
V_{\gamma'}(g)\Omega'\|\le\|\tilde V(\gamma'=1,g)\Omega'\|
\le\|g\|_2=1\eqno(68)$$ and finally $$|(\Psi,\tilde
V_\gamma(f)\Phi)|\le\|\Psi\|\|\Phi\|\|f\|_2\eqno(69)$$ for arbitrary $\Psi$ and
$\Phi$ in $\tilde {\cal F}_\gamma$. This proves boudedness of $\tilde
V_\gamma(f)$ with norm smaller or equal $\|f\|_2$, i.e. the same bound as for
$\gamma=1$. Since $\gamma,\gamma'\in {\bf R}$, $\gamma^2+\gamma^{\prime 2}=1$,
we have obtained boundedness of vertex operators for $\gamma\in [-1,1]$.

Note that in (61) we could have taken $\gamma,\gamma'$ complex because all
computations in the Heisenberg algebra do not depend on the involution. For
example, for computing $\|V_\gamma(z)\Phi_0\|^2$, $|z|<1$ we write
$$\|V_\gamma(z)\Phi_0\|^2=(V_\gamma(z)\Phi_0,V_\gamma(z)\Phi_0),$$ expand the
exponentials in $V_\gamma(z)$ and use the commutation relations, only. The usual
computation over the formal adjoint
$$(V_\gamma(z)\Phi_0,V_\gamma(z)\Phi_0)=(\Phi_0,V_\gamma(z)^+V_\gamma(z)
\Phi_0)$$ and the CBH formula is convenient [16] but not compulsory. It is the
fact that we have to realize $A_n$ in (61) as Hilbert space operators with
involution $A_n^+=A_{-n}$ that forces $\gamma,\gamma'$ to be real.  Hence, until
now we have proved the result for $\gamma\in [-1,1]$ only.

We now extend the region to the closed unit disc $|\gamma|\le 1$. Indeed the
vacuum correlation functions of vertices depend only on the absolute value
$|\gamma|^2$ and this proves the extension. Let us summarize the result of this
and the last section in the following

{\bf Theorem 3} The chiral vertex operators $\tilde V_\gamma(f)$, $f\in
L^2(S^1)$ smeared on the unit circle are bounded operators in the Hilbert space
$\tilde {\cal F}_\gamma^\oplus$ for all $\gamma\in {\bf C}$ with $|\gamma|\le
1$.

For similar results concerning $V_\gamma(f)$ see next section.  It is possible
that this result could be true for other values of $\gamma$, too, but other
techniques should be used. In the next section we would like to comment on the
results obtained in this paper.

\section{Remarks and discussion}

We have proved that contrary to the chiral unsmeared vertices, the smeared ones
are well behaved, being bounded operators in the Hilbert space $\tilde{\cal
F}^\oplus$ constructed explicitly in section 4 by starting from the bosonic Fock
space in which the vertices naturally live. The smearing in the complex variable
$z$ is only one-dimensional on the unit circle $|z|=1$, this is typical for free
fields in quantum field theory.

Regarding the Hilbert space, there are further possibilities to realize it.  We
give one example (other constructions can be found in [17]). Consider the vertex
operator $V(z)$ as in (3) with $\gamma=1$. Let $W$ be a finite linear
combination of vectors $V_n(f_j)$ , $f_j=(f_{j1},\ldots, f_{jn})$:
$$W=\sum_nV_n=\sum_{n,j}\alpha_jV_n(f_j)\eqno(70)$$ where in $V_n$ we collect
all terms in $W$ with the same number $n$ of vertex factors $V$. In the bosonic
Fock space ${\cal F}$ we change the scalar product $(\cdot,\cdot)$ to a new one
which we define on vectors $W$ by linearity and relations $$s(V_n,V_m)=0,\quad
{\rm if}\quad n\ne m$$ $$s(V_n,V_n)=(V_n,V_n).\eqno(71)$$ Let us remark that the
set of vectors $W$ (74) is dense in ${\cal F}$ (see Appendix 2). We complete in
the new scalar product and get a Hilbert space $\hat{\cal F}$. In both ${\cal
F}$, $\hat{\cal F}$ the set of vectors $W$ is contained densely with respect to
the different scalar products involved. In $\hat{\cal F}$ we introduce a densely
defined bilinear form $\hat V(f_0)$, $f_0\in L^2(S^1)$ by $$\hat
V(f_0)(V_n,V_m)=0\quad{\rm for}\quad m\ne n-1$$ $$\hat
V(f_0)(V_n,V_{n-1})=(V_n,V(f_0)V_{n-1}).\eqno(72)$$ Again the generalized Gram
inequality followed by some elementary reasoning shows that $\hat V(f_0)$ is a
densely defined bounded bilinear form in  $\hat{\cal F}$ w.r.t. the new scalar
product $s(\cdot,\cdot)$.  In conclusion it defines a bounded operator $\hat
V(f)$ in $\hat{\cal F}$.

In this approach, as well as in sections 3,4 (where also the case $\gamma\ne 1$
was considered), the spaces $\hat{\cal F}$ and $\tilde{\cal F}$, although
showing fermionic flavors, were directly coined in the bosonic Fock space in
which vertices are formally defined.  Neutrality was essentially used (76). Up
to section 5 braiding was not introduced. In section 5 we used it although we
belive that this would not be necessary if we could replace the abstract
reasoning on fermions by a more direct computation in bosonic Fock space. We do
not know what happens with the smeared vertices without neutrality, but we hope
to return to this interesting problem elsewhere. Usually neutrality is motivated
by the zero mass limit in two-dimensional field theory [18, 19]. Our interest in
obtaining smeared vertices as bounded operators in a Hilbert space as close as
possible to the bosonic Fock space was stimulated at the beginning by the idea
of applying the causal Epstein-Glaser method [5] to the two-dimensional
sine-Gordon model as well as to conformal perturbation theory. These two aspects
of two-dimensional physics are strongly related [20]. Operatorial boundedness in
both chiral and antichiral variables in a Minkowski approach will enable us to
prove not only convergence of perturbation theory, which is an ancient result,
but also to save a factorial factor in the estimate, leading to entire type in
the result and convergence for all coupling constants. A similar (weaker) result
for the case of two-dimensional conformal quantum field theory was already
mentioned in [4]. We will come back to these questions elsewhere.

Last but not least sections 2 and 3 can be looked at as a first step towards a
Hilbert space interpretation of vertex operator algebras (see for instance
[28]). Indeed, through the smearing out operation the calculus of formal
distributions (formal Dirac function) can be taken over to the Fourier space.
Using truncation of formal expansions and going to the unit circle, the
convergence is garanteed independent of the side from which the unit circle is
approached. Passing from inside to outside or vice versa is an involution on
Fourier coefficients.  According to the Hilbert space framework of this paper
the calculus of formal distributions is then a formal calulus on kernels of
Hilbert space operators.

\appendix\section*{Appendix 1: Gram determinant inequalities}

Here we state some generalized Gram inequalities for the generalied Gram
determinant. By the generalized Gram determinant we understand $$G(x,y)\equiv
G_p(x_1,x_2,\ldots,x_p;y_1,y_2,\ldots,y_p)=$$
$$=\left\vert\matrix{(x_1,y_1)&\ldots&(x_1,y_p)\cr\ldots\cr (x_p,y_1)&
\ldots&(x_p,y_p)\cr}\right\vert\eqno(A.1)$$ where $x=(x_1,\ldots,x_p)$,
$y=(y_1,\ldots,y_p)$ are in the same vector space with scalar product
$(\cdot,\cdot)$. The usual Gram determinant is obtained by taking $x_i=y_i$,
$i=1,\ldots,p$.

Here we consider the case where $x_i,y_j$ are $L^2$ functions. Let $F$ and $G$
be matrix functions given by $$F=\pmatrix{f_1\cr f_2\cr\ldots\cr
f_r\cr}\equiv\pmatrix{f_{11}&\ldots &f_{1n}\cr f_{21}&\ldots&f_{2n}\cr\ldots\cr
f_{r1}&\ldots&f_{rn}\cr} \eqno(A.2)$$ $$G=\pmatrix{g_1\cr g_2\cr\ldots\cr
g_s\cr}\equiv\pmatrix{g_{11}&\ldots &g_{1m}\cr g_{21}&\ldots&g_{2m}\cr\ldots\cr
g_{s1}&\ldots&g_{sm}\cr}.  \eqno(A.3)$$ We call the attentioan of the reader to
the fact that here $f_i$, $g_k$ represent vector functions. The generalized Gram
inequality we use in this paper is $$\sum_{{1\le j,j'\le r}\atop{1\le k,k'\le
s}}\alpha_j^*\beta_k^*\alpha _{j'}\beta_{k'}G_{n+m}(f_j,g_k;f_{j'}g_{k'})\le$$
$$\le\sum_{1\le j,j'\le r}\alpha_j^*\alpha_{j'}G_n(f_j;f_{j'}) \sum_{1\le
k,k'\le s}\beta_k^*\beta_{k'}G_m(g_k;g_{k'}),\eqno(A.4)$$ where $\alpha_j$,
$j=1,\ldots,r$; $\beta_k$, $k=1,\ldots,s$ are complex constants. The inequality
used in (50) is the special case $n=r=s=1$, $\alpha_1=1$ and the inequality (54)
is the case $n=r=1$, $\alpha_1=1$ ($\alpha$'s in (54) are $\beta$'s in (A.4)).
All inequalities for the usual Gram determinants [21] are special cases of (A.4)
with identical $f$ and $g$ functions.

Concerning the proof one could use the Landsberg formula [22]
$$\int\limits_{\lap^p}\left\vert\matrix{x_1(s_1)&\ldots&x_1(s_p)\cr\ldots\cr
x_p(s_1)&\ldots&x_p(s_p)\cr}\right\vert\left\vert\matrix{y_1(s_1)&\ldots
&y_1(s_p)\cr\ldots\cr y_p(s_1)&\ldots&y_p(s_p)\cr}\right\vert\,d_p\nu(s)=$$
$$=p!\,\left\vert\matrix{(x_1,y_1)&\ldots&(x_1,y_p)\cr\ldots\cr
(x_p,y_1)&\ldots&(x_p,y_p)\cr}\right\vert=p!G(x,y)\eqno(A.5)$$ where $x_i,y_j\in
L^2(\lap,\nu)$, $\lap^p=\lap\times\ldots\times\lap$ and $d_p\nu$ is the product
measure, together with Laplace expansion formula for determinants.

Certainly another proof can be given by fermionic Fock space methods. Consider
(smeared) fermionic annihilation and creation operators $a(f)$, $a^+(f)$:
$$\{a(f),a(g)\}=0=\{a^+(f),a^+(g)\}$$ $$\{a(f),a^+(g)\}=(f,g),\eqno(A.6)$$ where
$\{\cdot,\cdot\}$ is the anticommutator. The action in the fermionic Fock space
is as usual given by $$(a(f)\psi)^{(n)}(x_1,\ldots,x_n)=\sqrt{n+1}\int
dx\,f(x)^*\psi^{(n+1)} (x,x_1,\ldots,x_n)\eqno(A.7)$$
$$(a^+(f)\psi)^{(n)}(x_1,\ldots,x_n)=\sqrt{n}\sum_{i=1}^n(-1)^{i-1}f(x_i)
\psi^{(n-1)}(x_1,\ldots,\hat x_i,\ldots x_n)\eqno(A.8)$$ where $\hat x_i$
indicates that the $i$-th variable is to be omitted and
$\psi^{(n)}(x_1,\ldots,x_n)$ is totally antisymmetric.

Let $$\Psi=\sum_{j=1}^r\alpha_ja^+(f_{j1})\ldots a^+(f_{jn})\eqno(A.9)$$
$$\Phi=\sum_{k=1}^s\beta_k a^+(g_{k1})\ldots a^+(g_{km}).\eqno(A.10)$$ Then we
have on the vacuum $\Omega$
$$\|\Psi\Omega\|^2=\sum_{j,j'}\alpha_j^*\alpha_{j'}G_n(f_j;f_{j'}) \eqno(A.11)$$
$$\|\Phi\Omega\|^2=\sum_{k,k'}\beta_k^*\beta_{k'}G_m(g_k;g_{k'}) \eqno(A.12)$$
and $$\|\Psi\Phi\Omega\|^2=\sum_{j,j',k,k'}\alpha_j^*\beta_k^*\alpha_{j'}
\beta_{k'}G_{n+m}(f_j,g_k;f_{j'},g_{k'}).\eqno(A.13)$$ Hence, the generalized
Gram inequality (A.4) is proved if we can show that the operator norm
$\|\Psi\|^2$ is equal to
$$\|\Psi\|^2=\sum_{j,j'}\alpha_j^*\alpha_{j'}G_n(f_j;f_{j'}).  \eqno(A.14)$$

This is a consequence of Wick's theorem about normal ordering of operator
products which we write down with the following simplified notation
$$a(f_n)\ldots a(f_1)a^+(g_1)\ldots a^+(g_n)=\>:a(f_n)\ldots a^+(g_n):$$
$$+:\kon{a(f_n)\ldots}{a^+}(g_j)\ldots a^+(g_n):+\ldots +G(f_1,\ldots,
f_n;g_1,\ldots,g_n).\eqno(A.15)$$ The r.h.s. is obtained from the l.h.s. by
normal ordering denoted by double dots, that means by anticommuting all emission
operators $a^+$ to the left. The "contractions" (indicated by the bracket
upstairs) represent the anticomutators $(f_n,g_j)$ (A.6) which appear in this
process. The last term has all operators contracted in pairs in all possible
ways and this gives just the Gram determinant.

Now let us consider the square
$$(\Psi\Psi^+)^2=\Bigl[\sum_{j,j'}\alpha_j\alpha_{j'}a^+(f_{j1})\ldots
a^+(f_{jn})a(f_{j'n})\ldots a(f_{j'1})\Bigl]^2$$
$$=\sum_{jj'll'}\alpha_j\alpha_{j'}^*\alpha_l\alpha_{l'}^*a^+(f_{j1}) \ldots
a^+(f_{jn})\times\eqno(A.16)$$ $$\times\>a(f_{j'n})\ldots
a(f_{j'1})a^+(f_{l1})\ldots a^+(f_{ln})a(f_{l'n} \ldots a(f_{l'1}).$$ In the
last line we substitute Wick's theorem (A.15). Then only the last term with
Gram's determinant contributes because all other terms contain at least two
equal Fermi operators $a(f)a(f)=0$. This gives
$$(\Psi\Psi^+)^2=\sum_{j',l}\alpha_{j'}^*\alpha_lG(f_{j'1},\ldots
f_{j'n};f_{l1},\ldots f_{ln})$$
$$\times\sum_{j,l'}\alpha_j\alpha_{l'}^*a^+(f_{j1})\ldots a^+(f_{jn})
a(f_{l'n})\ldots a(f_{l'1})=$$ $$=(\sum\alpha^*\alpha G)(\Psi\Psi^+)$$ with
obvious short-hand notation.  Since $\Psi\Psi^+$ is selfadjoint this implies
$$\|\Psi\Psi^+\|=|\sum\alpha^*\alpha G|=\|\Psi\|^2=\|\Psi^+\|^2$$ which is the
desired result (A.14).

\appendix\section*{Appendix 2}

Strictly speaking the result of this appendix is not necessary for the main body
of the paper. This is because we realized neutrality by passing from the genuine
bosonic Fock space ${\cal F}$ to Hilbert spaces $\tilde {\cal F}$ (see sections
3-5) where linear combinations of vectors $\tilde V_n$ are dense by
construction. The same applies to the space $\hat{\cal F}$ in section 6.
Nevertheless the results of this appendix help to get a better understanding of
the structure of the Hilbert spaces in which our vertices live. It could gain on
interest if we succeed to abolish neutrality. Furthermore, the problem touched
here was mentioned as open in a related setting up in [24] (see also [27]).

It is our aim to show that the set of vectors
$$V_n(f)=\prod_{i=1}^nV(f_i)\Phi_0\eqno(A.20)$$ where $f=(f_1,\ldots,f_n)$,
$f_i\in L^2(S^1)$ is total in the Fock space ${\cal F}$. In other words linear
combinations $$\sum^{n,j}\alpha_jV_n(f_j),\quad\alpha_j\in{\bf C},\quad
n=0,1,2,\ldots\eqno(A.21)$$ where now $f_j=(f_{1j},\ldots,f_{nj})$ are vector
functions dense in ${\cal F}$. This type of property is common in quantum field
theory and we could invoke this. The point is that vertices $V(f)$ are not
exactly quantum fields because of the massless limit involved. A direct clean
proof  of this property is therefore desirable. This point was already mentioned
in [24]. It turns out that the vectors (A.20) are coherent states of a special
type. One expects coherent states to be dense (even overcomplete [14]).

We use the relation for the unsmeared vertices $$\prod_{i=1}^nV(z_i)=\prod_{1\le
i<j\le n}\Bigl(1-{z_j\over z_i}\Bigl) \exp\biggl(\sum_{i=1}^\infty{1\over
i}\sum_{j=1}^nz_j^ia_{-i}\biggl)$$ $$\times\exp\biggl(\sum_{i=1}^\infty{1\over
i}\sum_{j=1}^nz_j^{-i}a_i\biggl) $$ where $1>|z_1|>|z_2|>\ldots |z_n|$, and
apply it to the vacuum $\Phi_0$: $$\prod_{i=1}^nV(z_i)\Phi_0=\prod_{1\le i<j\le
n}\Bigl(1-{z_j\over z_i}\Bigl) \exp\biggl(\sum_{i=1}^\infty{1\over
i}Z_i^{n}a_{-i}\biggl)\Phi_0\eqno(A.22)$$ with $Z_i^{(n)}=\sum_{j=1}^n z_j^i$.
Now $$\exp\biggl(\sum_{i=1}^\infty{1\over i}Z_i^{n}a_{-i}\biggl)\Phi_0
\eqno(A.23)$$ are typical coherent states [14]. The idea is to integrate (A.22)
against powers of $z_i^{-1}$ and use $${1\over 2\pi
i}\int_{S^1}z^ndz=\delta_{n,-1}$$ in order to produce elements $\Phi_\eta$ of
the Fock basis out of (A.22). For instance, if $n=1$, $Z_i=z_1^i$, we get
$$V(z_1^{-2})\Phi_0={1\over 2\pi}\int V(z_1)z_1^{-2}dz_1\Phi_0$$
$$\sim\int\Bigl(1+{a_{-1}\over 1}z_1+{a_{-2}\over 2}z_1^2+\ldots\Bigl)
z_1^{-2}dz_1\Phi_0=$$ $$=a_{-1}\Phi_0\sim\Phi_\eta,\quad\eta=(1,0,0,\ldots).$$
How to obtain states like $a_{-2}\Phi_0$ and $a_{-1}^2\Phi_0$ ?

Let us expand $$V(z_1)\Phi_0=\exp\sum_{n=1}^\infty{z_1^n\over n}a_{-n}\Phi_0$$
$$=\biggl[1+\sum{z_1^n\over n}a_{-n}+{1\over 2!}\Bigl(\sum{z_1^n\over
n}a_{-n}\Bigl)^2+\ldots\biggl]\Phi_0$$ and collect the terms with $z_1^n$:
$${a_{-n}\over n}+{1\over 2!}\sum_{n_1+n_2=n}{a_{-n_1}a_{-n_2}\over
n_1n_2}+{1\over 3!}\sum_{\sum n_j=n}{a_{-n_1}a_{-n_2}a_{-n_3}\over
n_1n_2n_3}+\ldots\eqno(A.24)$$ The corresponding Fock state is obtained by
integration with $z_1^{-n-1}$. We now consider two factors and use (A.22)
$$V(z_1)V(z_2)\Phi_0=g(z_1,z_2)\exp\sum_{n=1}^\infty{z_1^n+z_2^n \over
n}a_{-n}\Phi_0$$ $$=g(z_1,z_2)\biggl(\exp\sum_{n=1}^\infty{z_1^n\over
n}a_{-n}\biggl) \biggl(\exp\sum_{n=1}^\infty{z_2^n\over
n}a_{-n}\biggl)\Phi_0,\eqno(A.25)$$ where $g(z_1,z_2)$ are the prefactors in
(A.22). The latter are cancelled by working with the following smearing function
$g^{-1}z_1^{-n-1}z_2^{-m-1}$. Since we can choose $n$ and $m$ independently,
this gives us arbitrary products of two factors of the form (A.24). The
generalization to more than two factors is straight-forward.

It is not hard to see that in this way we get a total set of Fock vectors.
Indeed, starting with $n=1$ in (A.24) gives $a_{-1}\Phi_0$, and the
$\eta_1$-fold product with itself (denoted by $1^{\eta_1}$) gives
$a_{-1}^{\eta_1}\Phi_0$. Next we take $n=2$ in (A.24): $${a_{-2}\over 2}+{1\over
2!}a_{-1}^2.$$ Since $a_{-1}^2\Phi_0$ is already constructed, we get
$a_{-2}\Phi_0$.  Forming products $1^{\eta_2}\times 2$, we find
$a_{-1}^{\eta_1}a_{-2} \Phi_0$. Now we are ready to consider the product $(2^2)$
of two $n=2$: $${1\over 4}(a_{-2}^2+2a_{-2}a_{-1}^2+a_{-1}^4).$$ This gives us
$a_{-2}^2\Phi_0$ because the other two vectors are already known. The next steps
in the process are:
$$a_{-1}^{\eta_1}a_{-2}^2\Phi_0,\>a_{-2}^3\Phi_0,\>a_{-1}^{\eta_1}
a_{-2}^3\Phi_0,\ldots,a_{-2}^{\eta_2}a_{-1}^{\eta_1}\Phi_0.$$ We continue with
$n=3$ and so on. We thus arrive at the usual basis in Fock space. The same proof
goes through for $\gamma\ne 1$.

The test functions we have used in the foregoing construction are of a more
general kind than the simple products of $f_i(z_i)$ in (A.20). But this is no
essential point because such test functions can be approximated by linear
combinations of simple products with help of the Laurent expansion.

In particular the above proof shows that the set of smeared coherent states
$$\prod_{i=1}^n V_-(f_i)\Phi_0,\quad n=0,1,2,\ldots\eqno(A.26)$$ where $f_i\in
L^2(S^1)$ is total in ${\cal F}$. Concerning the unsmeared case, the situation
is not completely clear to us. We do not know whether the set (A.26) where the
smearing is left out and the variables satisfy $1>|z_1|>|z_2|\ldots >|z_n|$ is
total in ${\cal F}$. On the other hand a slightly larger set based on
$$\exp\Bigl(\sum_n{t_n\over n}a_{-n}\Bigl)\Phi_0$$ where now $t_n$ are left
arbitrary in ${\bf C}$ is total [11]. These are the genuine coherent states in
infinitely many variables.  \vskip 0.5cm {\bf Acknowledgements} \vskip 0.5cm We
thank W.Boenkost for assistance at an incipient stage of this work, K.-H.Rehren
for sending published and unpublished work and F.Kleespiep for discussions. One
of us (F.C.) thanks Giovanni Felder and Forschungsinstitut f\"ur Mathematik, ETH
Z\"urich for hospitality and to J\"urg Fr\"ohlich for encouragement.

 \end{document}